\documentclass[aps,showpacs]{revtex4}

\usepackage{graphicx}
\usepackage{dcolumn}
\usepackage{amsmath}

\newcommand
 {\pAu}       {\mbox{p-Au}}
\newcommand {\piAu}      {\mbox{$\pi$-Au}}
\newcommand{\pbarAu}  {\mbox{$\bar{\text{p}}$-Au}}

\begin{document}

\title[Short Title]{Centrality dependence of the thermal excitation-energy deposition in 8--15~GeV/c hadron-Au reactions} 

\author{R.A. Soltz}\email{soltz@llnl.gov}
\author{R.J. Newby}
\author{J.L. Klay}\altaffiliation[Present address:]{California Polytechnic State University, San Luis Obispo, CA 93407}
\author{M. Heffner}
\affiliation{N-Division, Lawrence Livermore National Laboratory, 7000 East Avenue, Livermore, CA 94550, USA}

\author{L. Beaulieu}\altaffiliation[Present address:]{Laval University, Quebec City, Quebec, Canada G1K7P4}
\author{T. Lefort}\altaffiliation[Present address:]{LPC de Caen, 14050 Caen cedex, France}
\author{K. Kwiatkowski}\altaffiliation[Present address:]{Los Alamos National Laboratory, Los Alamos, New Mexico 87545}
\author{V.E. Viola}
\affiliation{Department of Chemistry and IUCF, Indiana University, Bloomington, Indiana 47304}

\author{for the E900 Collaboration}

\date{\today}

\begin{abstract}
The excitation energy per residue nucleon (E*/A) and fast and thermal light particle multiplicities are studied as a function of centrality defined as the number of grey tracks emitted $N_{\rm grey}$ and by the mean number of primary hadron-nucleon scatterings ($\left<\nu\right>$) and mean impact parameter ($\left<b\right>$) extracted from it.  The value of E*/A and the multiplicities show an increase with centrality for all systems, 14.6~GeV \pAu~and 8.0~GeV \piAu~and \pbarAu~collisions, and the excitation energy per residue nucleon exhibits a uniform dependence on $N_{\rm grey}$.
\end{abstract}


\maketitle

\section{Introduction}
\label{sec:intro}


Multifragmentation is the process by which a heavy nucleus excited through nuclear collision decays into nuclear fragments with a range of atomic masses and energies.  First observed in cosmic ray emulsion experiments as a starburst of grey and black emulsion tracks emanating from a single incident track, it has since been studied using beams of projectiles including single hadrons (pions, protons, and antiprotons), light ions, and heavy ions accelerated to energies ranging from a few hundred MeV up to Tevatron energies of 350~GeV~\cite{Denisenko:1958un,Hauger:1998jg,Hirsch:1984yj}.  

Analysis of the hadron-induced reactions suggests a fast evolutionary mechanism that is dominated by two stages~\cite{Viola:2006fh,Beaulieu:2001pc}. Initially the system is heated by a prompt nuclear cascade that produces high-energy, forward peaked particles, commonly identified as grey particles, named according to their appearance in emulsion. The system then cools via preequilibrium particle emission, leading to the second major stage, an excited, thermalized nuclear residue that subsequently undergoes statistical decay.  Analyses of the second stage emission products indicate that for events that deposit the highest excitation energies, the residue undergoes multifragment decay on a near-instantaneous time scale~\cite{Beaulieu:2000ci}, consistent with a nuclear liquid-gas phase transition and perhaps critical behavior~\cite{KleineBerkenbusch:2001kq,Elliott:2001hn}.  

The process of multifragmentation has been thoroughly investigated as a function of incident beam energy~\cite{Hirsch:1984yj} and projectile species~\cite{Viola:2006fh}, but a systematic study of its centrality dependence has not yet been performed.  In heavy ion collisions, the experimental definition of centrality usually refers to the mean number of primary collisions or the impact parameter distribution extracted from forward particle multiplicities using a standardized set of assumptions, and it is often used to characterize nuclear collisions at GeV energies and above~\cite{Chemakin:2000ha,Adler:2007by}.  Centrality analyses often provide useful insights into particle production mechanisms and enable systematic comparisons with other experiments and theoretical models, where comparable measures of centrality can be precisely determined.  The ISiS multifragmentation program at the Brookhaven National Laboratory Alternating Gradient Synchrotron (BNL--AGS) with its large acceptance and event-by-event charged particle identification is ideally suited to study the centrality dependence of multifragmentation.  In this paper we present results from the first multifragmentation centrality analysis using the data from the ISiS experiments E900 and E900a.

\section{Experiment}
\label{sec:exp}
The experimental grey particle and excitation energy distributions were obtained at the AGS using secondary beams of untagged 14.6 GeV/c protons and tagged beams of 8.0 GeV/c $\pi^-$ and antiprotons incident on a $^{197}$Au target. The highest energy data sets for each beam species were selected to optimize the Glauber model assumptions required for the centrality analysis.
Reaction products were measured with the ISiS detector array, which consists of 162 triple detector telescopes arranged in a spherical geometry and covering 74\% of 4$\pi$~\cite{Kwiatkowski:1995ye}. Detector telescopes were composed of a low-pressure gas ionization chamber, followed by a 500~$\mu$m silicon detector for measuring low-energy fragments and a 28~mm CsI crystal for detecting more energetic particles. Telescopes provided charge identification for 1.0$<$E/A$<$90~MeV reaction products and particle identification in the range  8.0$<$E/A$<$92~MeV, primarily hydrogen and helium isotopes. 

Of particular relevance to the present study, the kinetic energy of particles that punched through the CsI crystal was derived from the CsI energy loss in the Si-CsI particle-identification spectrum and energy-loss tables. This procedure permitted identification of particles up to 350 MeV in energy, which were assumed to be primarily hydrogen nuclei and were classified as grey particles. Full description of the experimental details, data analysis, calorimetry procedures and error estimates are given in~\cite{Lefort:2001pa,Viola:2006fh}.     

\section{Analysis}
\label{sec:ana}

\begin{figure}
\begin{center}
\includegraphics[width=0.65\textwidth]{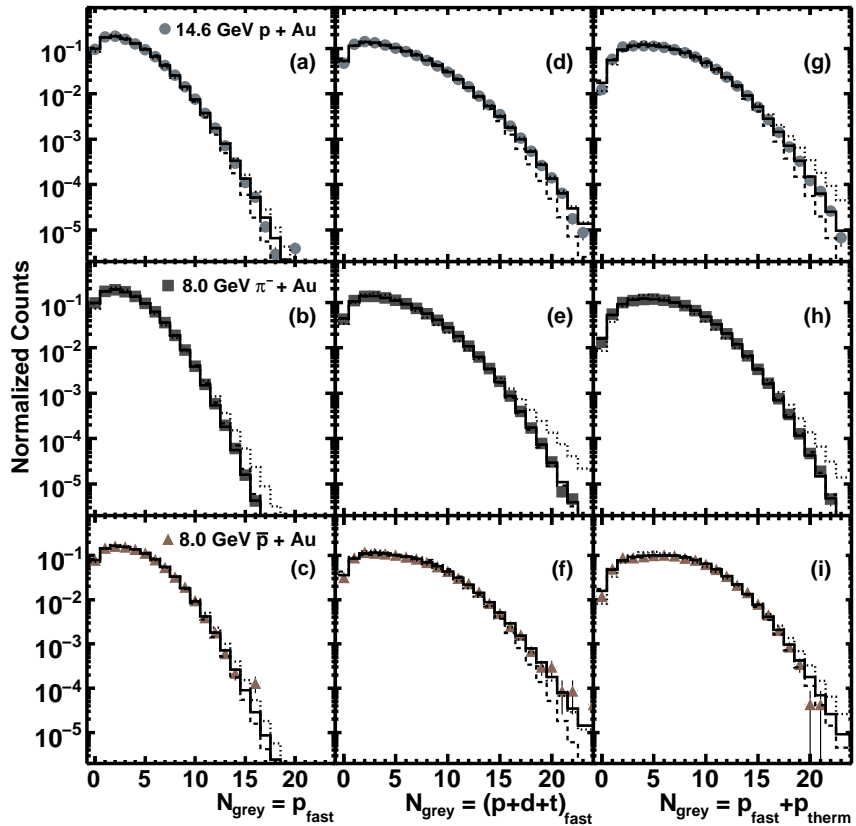}
\caption{(Color online) Fits to grey track distributions for incident protons are shown at the top, negative pions below, and antiprotons at the bottom.  Fits are shown for $N_{\rm grey}$ distributions with three definitions: fast protons (a,b,c), fast protons, deutrons and tritons (d,e,f) and fast and thermal protons (h,i,j).  Each distribution was fit to three functional forms, Eq.~\ref{eq:SS2} (solid), Eq.~\ref{eq:SS2} with $c_0=1$ (dashed), and a second order polynomial (dotted).}
\label{fig:DataSys}
\end{center}
\end{figure}

We follow the procedure developed by BNL-E910~\cite{Chemakin:1999jd,Hegab:1981iq,Andersson:1977wg}, in which the conditional probability for the grey track distribution as function of the number of hadron-nucleon scatterings ($\nu$), $P(N_{\rm grey}|\nu)$, was determined by a fit  to the measured grey particle distributions.  Here we use a modified exponential form for $P(N_{\rm grey}|\nu)$,
\begin{equation}
\overline{N_{\rm grey}}(\nu) = c_1 \left[ 1 -  \left( \exp(-\nu/c_{2}) \right) ^{c_0} \right].
\label{eq:SS2} 
\end{equation}
and assume a binomial distribution for the number of grey tracks given $Z$ total protons.  Eq.~\ref{eq:SS2} preserves the dominant linear term of the second order polynomial used in~\cite{Chemakin:1999jd}, but accommodates the $N_{\rm grey}$ saturation in a more natural way.  $N_{\rm grey}$ was defined as the number of identified protons with 30$<$E$<$350~MeV.  This energy range is meant to separate fast and thermal protons, and is consistent with previous definitions of $N_{\rm grey}$.  The lower energy cut is identical to the one used by BNL-E910, although their high energy cut of 585~MeV is higher than that used by most experiments.  The dependence on the low energy cut is included in the study of systematic errors.  The full functional form is given by Eq.~\ref{eq:binom},
\begin{eqnarray}
\label{eq:binom}
P(N_{\rm grey}) & = & \sum_{\nu} P(N_{\rm grey}|\nu) \pi(\nu), \\
P(N_{\rm grey}|\nu) & = & \left( \begin{array}{c} Z \\ N_{\rm grey} \end{array}  \right) 
X^{N_{\rm grey}} \left(1 - X \right) ^{Z-N_{\rm grey}}, \nonumber \\
X & = & \frac{\overline{N_{\rm grey}}(\nu)}{Z}. \nonumber
\end{eqnarray}
The $\pi(\nu)$ Glauber distribution for the number of primary hadron-nucleon interactions is calculated with a Monte Carlo optical model~\cite{Glauber:1970} using a Wood-Saxon nuclear density profile assuming free inelastic cross-sections for protons at 14.6 GeV/c and pions and antiprotons at 8.0~GeV/c, given in the particle data book as 30, 20, and 47~mb, respectively~\cite{pdg:2006}.  The Glauber distribution for the E900 data set includes a 10\% admixture of 14.6~GeV/c pions~\cite{Lefort:2001pa}, also using a 20~mb cross-section.  Fits of Eq.~\ref{eq:binom} to the fast proton $N_{\rm grey}$ distributions are shown as solid lines in Fig.~\ref{fig:DataSys} panels (a), (b), and (c).  The $\chi^2/ndf$ obtained for the proton, pion, and antiproton systems were 30.2, 11.9, and 1.0 respectively. To estimate systematic errors, fits were also performed with a version of Eq.~\ref{eq:binom} with $c_0$ set to unity (dashed) and to the second order polynomial (dotted) used in~\cite{Chemakin:1999jd}.  The $\chi^2/ndf$ values for these fits were orders of magnitude larger for some systems.  In a study of systematic errors reported in~\cite{Chemakin:1999jd} in which the $N_{\rm grey}$ analysis was compared to a microscopic cascade RQMD model~\cite{Sorge:1992} the overall systematic error was estimated to be 10--20\%, with the dominant contribution coming from the lower momentum bound in the $N_{\rm grey}$ definition.  For this reason we also performed fits to the combined fast and thermal proton distributions 8$<$E$<$350~MeV, shown in the Fig.~\ref{fig:DataSys} panels (g),(h), and (i), and to the sum of the fast proton, deuteron, and triton distributions, shown in the Fig.~\ref{fig:DataSys} panels (d), (e), and (f).  Note for the fast deuteron and triton definitions, the low energy cut is set to 49~MeV rather than 30~MeV used for protons.  The systematic error estimate also includes fits with cross-sections varied by $\pm 2$~mb.  The values of $\left<\nu\right>$ and $\left<b\right>$ are given by,
\begin{eqnarray}
\overline{\nu}(N_{\rm grey}) & = & \sum_{\nu} \nu P(N_{\rm grey}|\nu) \pi(\nu), \\
 \overline{b}(N_{\rm grey}) & = & \sum_{\nu,b} b \, P(N_{\rm grey}|\nu) \pi(\nu,b).
\label{eq:nub} 
\end{eqnarray}
The excitation energy per residue nucleon (E*/A) was determined by  the method described in~\cite{Lefort:2001pa} by summing the energy measured in all charged particle fragments with a parameterized contribution from neutrons based upon~\cite{Goldenbaum:1996ih}.  The thermal energy cuts used to calculate contributions to the excitation energy were the same, i.e. 30~MeV for protons and 49~MeV for deuterons and tritons.  Therefore, {\em E*/A is independent of the $N_{\rm grey}$ used to determine centrality} and for all of the systematic error evaluations except for the set of fits that specifically include the thermal contribution.

\section{Results}
\label{sec:res}

\begin{figure}
\begin{center}
\includegraphics[width=0.49\textwidth]{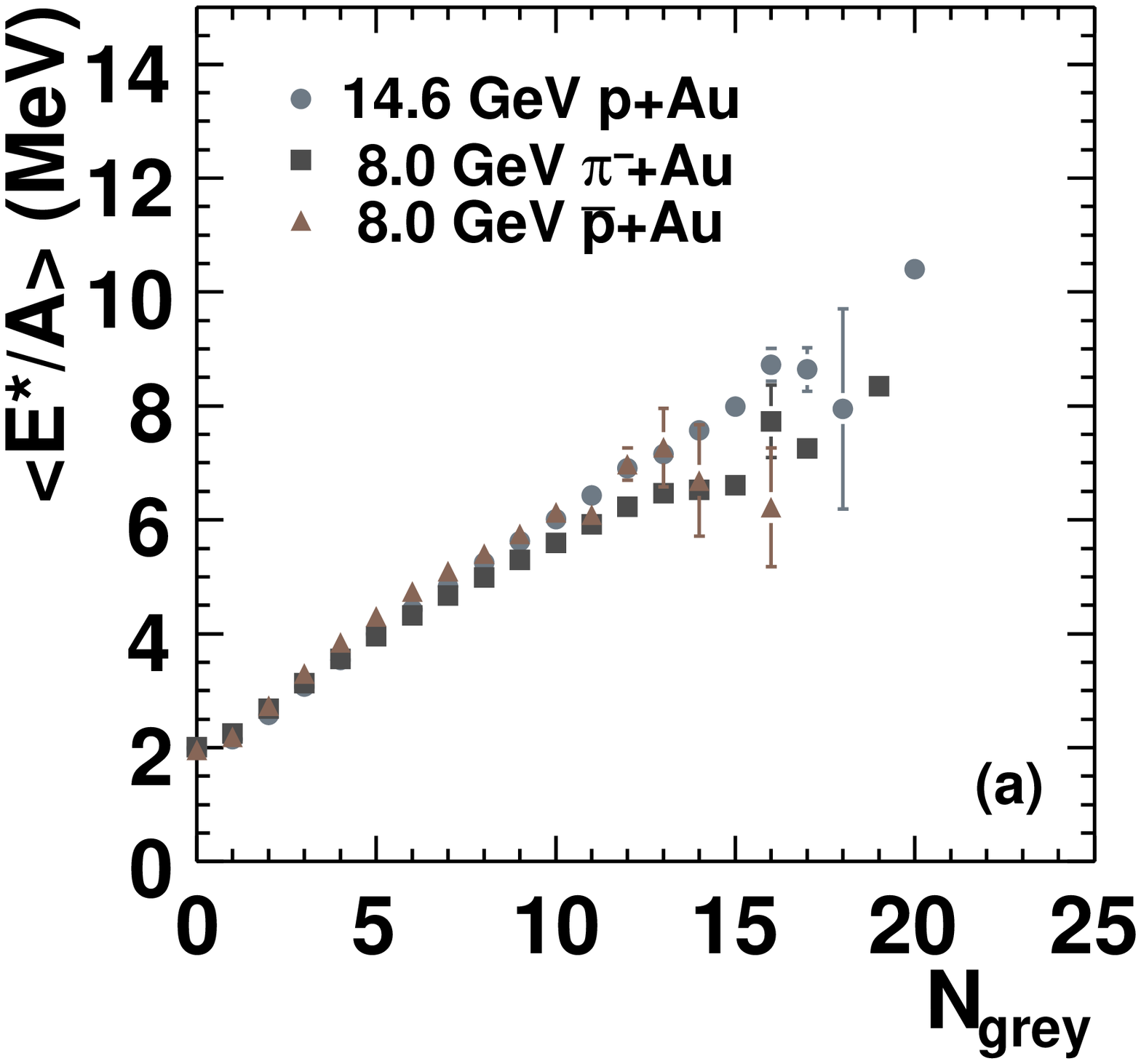}
\includegraphics[width=0.49\textwidth]{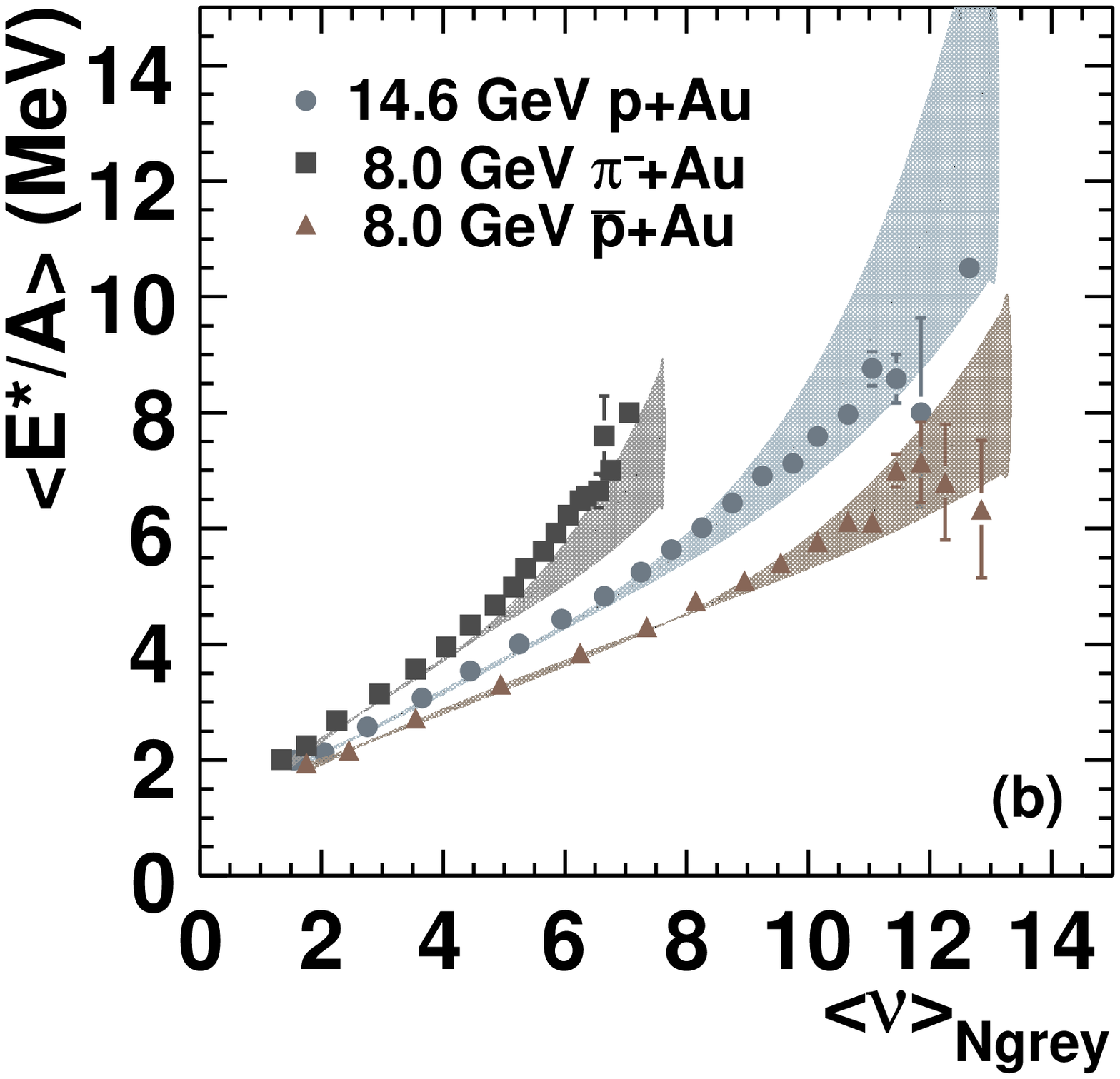}
\includegraphics[width=0.49 \textwidth]{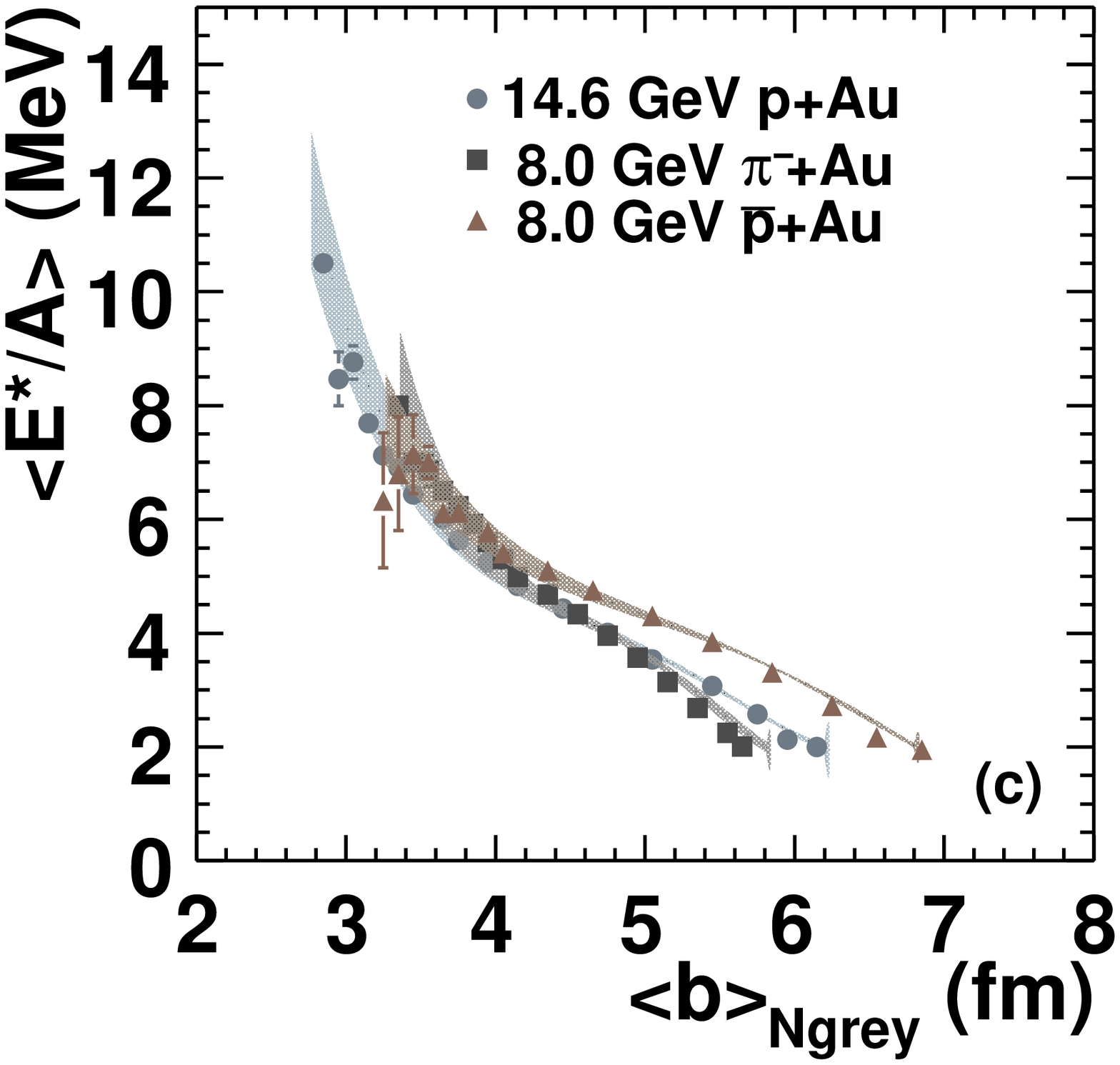}
\caption{(Color online) Excitation energy E*/A vs. $N_{\rm grey}$ (a), the mean number of hadron-nucleon scatterings, $\left<\nu\right>$ (b), and the mean impact parameter, $\left<b\right>$ (c) plotted for three systems: \pAu~(circles), \piAu~(squares), and \pbarAu~(triangles). Filled bands indicate systematic errors.}
\label{fig:EoAnub}
\end{center}
\end{figure}
Fig.~\ref{fig:EoAnub} shows the excitation energy per residue nucleon as a function of the measured $N_{\rm grey}$ (a), and the extracted quantities for $\left<\nu\right>$ (b) and $\left<b\right>$ (c).  The systematic error bands show the RMS variation in E*/A for the three functional forms, the inclusion of thermal protons and fast deuterons and tritons in the $N_{\rm grey}$ definitions, and the $\pm 2$~mb variations in cross-section described earlier.  The systematic error bands are centered about the mean for the set of all variations, leading to a slight displacement from the standard analysis values that is most prominent for the pions.

\begin{figure}
\begin{center}
\includegraphics[width=0.65\textwidth]{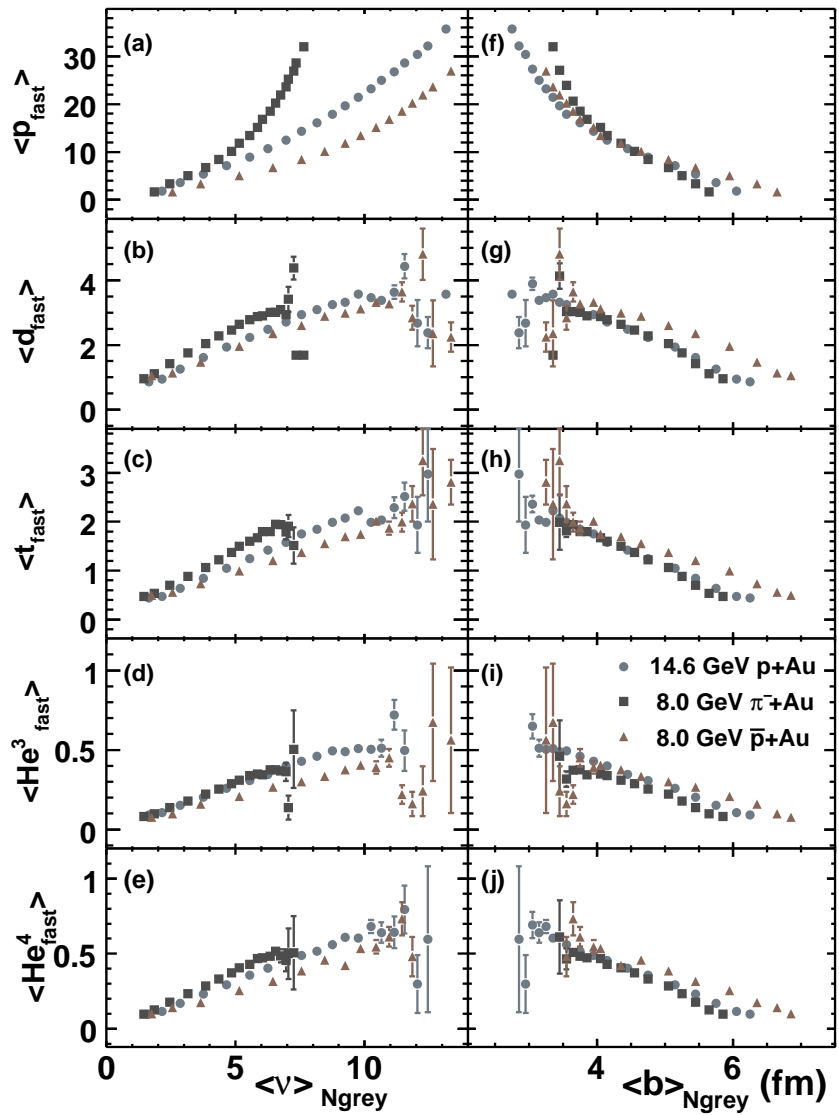}
\caption{(Color online) Mean fast particle multiplicities vs. mean number of hadron-nucleon scatterings ($\left<\nu\right>$) and mean impact parameter ($\left<b\right>$) extracted from the number of fast protons for three systems: \pAu (circle), \piAu (square), and \pbarAu (triangle)}.
\label{fig:MultFast}
\end{center}
\end{figure}

\begin{figure}
\begin{center}
\includegraphics[width=0.65\textwidth]{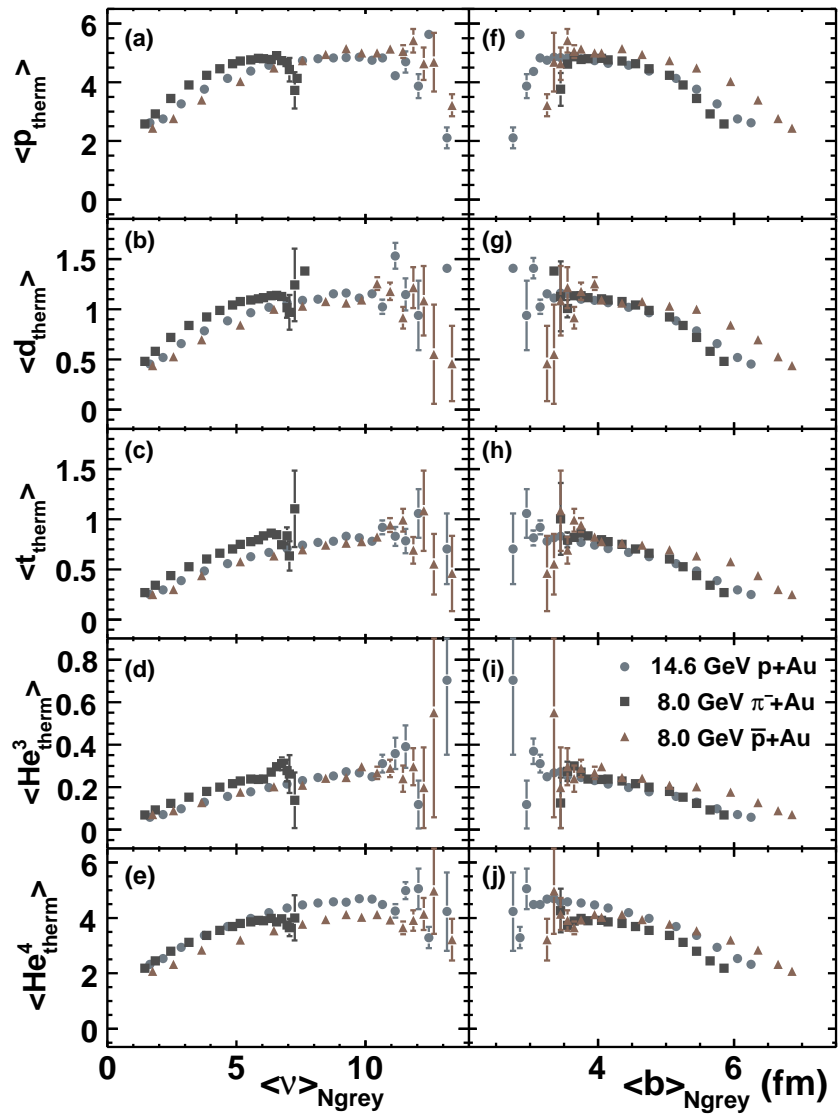}
\caption{(Color online) Mean thermal particle multiplicities vs. mean number of hadron-nucleon scatterings ($\left<\nu\right>$) and mean impact parameter ($\left<b\right>$) extracted from the number of fast protons for three systems: \pAu (circle), \piAu (square), and \pbarAu (triangle)}.
\label{fig:MultTherm}
\end{center}
\end{figure}

The dependence of E*/A on $N_{\rm grey}$ is approximately linear and nearly uniform for the three systems, with a slightly higher energy deposition for the antiprotons.  This uniformity does not extend to the extracted mean number of scatterings and mean impact parameter, except in the case impact parameter dependence for the most central collisions ($\left<b\right> < 4$~fm).  The relation between $N_{\rm grey}$ and the calculated values of $\left<\nu\right>$ and $\left<b\right>$ for each system is a direct consequence of the different hadron-nucleon cross-sections.  To reach the same $\left<\nu\right>$, the pions must traverse a much thicker region of the nucleus than the protons or antiprotons.  This implies a smaller impact parameter and leads to a larger number of fast protons emitted and therefore a greater excitation of the nucleus.   This is illustrated in Fig.~\ref{fig:MultFast} panels (a) and (f) which show the fast proton multiplicity ($N_{\rm grey}$) vs. the extracted quantities of $\left<\nu\right>$ and  $\left<b\right>$.  The remaining panels show the dependence of other light fragments.  Fig.~\ref{fig:MultTherm} shows the dependence of the thermal particle emission on the extracted centrality measures.
The mean fragment values  in Figures~\ref{fig:MultFast} and~\ref{fig:MultTherm} include efficiency corrections described in~\cite{Lefort:2001pa,Beaulieu:2001pc}.  Note that small differences for the most central collisions are not significant relative to the systematic errors, which have been omitted for clarity.

\section{Conclusions}
\label{sec:conc}

We have presented the first measurements of nuclear excitation energy and light charged particle emissions as a function of centrality.  As expected, the excitation energy increases with centrality (see Fig. 10 of~\cite{Turbide:2004ac}), but it is most directly related to the number of emitted grey particles defined as the fast protons with 30$<$E$<$350~MeV.  This is consistent with the two stage multifragmentation process, in which a prompt intranuclear cascade depletes and excites the remnant nucleus which then fragments and decays.  The different dependences on $\left<\nu\right>$ and $\left<b\right>$ serve to illustrate the importance of the collision geometry in understanding how the cascade develops from the initial primary collisions.  This analysis will serve as an important constraint for models such as the intranuclear cascade (INC)~\cite{Yariv:1979zz,Toneev:1990vj}, the hybrid statistical multifragmentation model (SMM)~\cite{Botvina:1990ke}, and the Boltzmann-Uehling-Uhlenbeck model (BUU)~\cite{Bertsch:1984gb}.  In particular, it will be interesting to see if the measured $N_{\rm grey}$ dependence can be reproduced, and whether the extracted trends in excitation energy and light fragment emission {\em vs.} $\left<\nu\right>$ and $\left<b\right>$ are borne out within the detailed models.

\section{Acknowledgements}
\label{sec:ack}
R.A.~Soltz wishes to thank the E900 collaboration in full and B.A.~Cole for initial discussions on the potential of the E900 data sets.  This work was performed under the auspices of the U.S. Department of Energy by Lawrence Livermore National Laboratory in part under Contract W-7405-Eng-48 and in part under Contract DE-AC52-07NA27344.

\end{document}